\documentclass[journal=nl,manuscript=article]{achemso}
\usepackage{amsmath}
\usepackage{graphicx}
\usepackage{epsfig}
\usepackage{dcolumn}
\usepackage{bm}
\usepackage{rotating}
\usepackage{multirow}
\usepackage{xspace}
\usepackage[table]{xcolor}
\usepackage[english]{babel}
\usepackage{colortbl}

\def\ve{{\varepsilon}}

\def\bk{{\bf k}}
\def\bG{{\bf G}}

\def\bq{{\bf q}}

\def\bt{{\bf t}}

\newcommand{\ket}[1]{ | #1 \rangle }
\newcommand{\bra}[1]{ \langle #1 | }

\def\>{\rangle}
\def\<{\langle}

\def\ok{$\overline{\rm K}$\xspace}
\def\oke{\overline{\rm K}}

\author{Fabio Caruso}
\email{caruso@physik.uni-kiel.de}
\affiliation{Institut f\"ur Theoretische Physik und Astrophysik, Christian-Albrechts-Universit\"at zu Kiel, Kiel, Germany} 
\author{Maximilian Schebek}
\affiliation{Institut f\"ur Physik and IRIS Adlershof, Humboldt-Universit\"at zu Berlin, Berlin, Germany} 
\author{Yiming Pan}
\affiliation{Institut f\"ur Theoretische Physik und Astrophysik, Christian-Albrechts-Universit\"at zu Kiel, Kiel, Germany} 
\author{Cecilia Vona}
\affiliation{Institut f\"ur Physik and IRIS Adlershof, Humboldt-Universit\"at zu Berlin, Berlin, Germany} 
\author{Claudia Draxl}
\affiliation{Institut f\"ur Physik and IRIS Adlershof, Humboldt-Universit\"at zu Berlin, Berlin, Germany} 

 \title{Chirality of Valley Excitons in Monolayer Transition-Metal Dichalcogenides} 

\begin{document}

\begin{abstract}
By enabling control of valley degrees of freedom in 
transition-metal dichalcogenides, valley-selective circular 
dichroism has become a key concept in valleytronics. 
In this manuscript, we show that valley excitons --  
bound electron-hole pairs formed at either the K or \ok 
valleys upon absorption of circularly-polarized light -- 
are chiral quasiparticles characterized by a finite orbital 
angular momentum (OAM). 
We further formulate an {\it ab-initio} many-body theory 
of valley-selective circular dichroism and valley excitons 
based on the Bethe-Salpeter equation. 
Beside governing the interaction with circularly polarized light, 
the OAM confers excitons a finite magnetization which manifests 
itself through an excitonic Zeeman splitting upon interaction with 
external magnetic fields. The good agreement between 
our {\it ab-initio} calculations and recent experimental 
measurements of the exciton Zeeman shifts corroborate this picture. 
\end{abstract}

\maketitle
\section{Introduction}

The two-fold valley degeneracy of monolayer transition-metal dichalcogenides (TMDs) 
makes them suitable candidates for the exploration of novel concepts in valleytronics \cite{rycerz_valley_2007}.
The rich valley physics of the TMDs manifests itself, for instance, in 
the formation of valley excitons \cite{yu_valley_2015}, 
chiral phonons \cite{zhu_observation_2018}, 
nonequilibrium phonon populations \cite{caruso2021},
as well as non-trivial topological properties \cite{yao_valley-dependent_2008},  
which are exemplified by the emergence of Hall effects in various flavors 
\cite{mak_valley_2014,qian_quantum_2014,onga_exciton_2017,dau_valley_2019}, 
and valley-dependent optical selection rules \cite{xiao_coupled_2012,PhysRevLett.120.087402}.

Circularly-polarized light can lead to a pronounced valley-selective 
circular dichroism (VCD) \cite{cao_valley-selective_2012,mak_control_2012,zeng_valley_2012,
jones_optical_2013,mak_tightly_2013,mak_photonics_2016,beyer_80_2019,PhysRevLett.125.216404}, 
whereby absorption is governed by the formation of bound electron-hole pairs (excitons) at either 
the K or \ok valleys in the Brillouin zone (BZ) 
depending on the light helicity. 
VCD enables to selectively tailor the population of excitons and excited carriers in the BZ, and it  
has opened new promising venues to achieve properties on demand using light \cite{basov_towards_2017}. 
Additionally, the exploitation of valley degrees of freedoms in TMDs 
relies on the possibility to establish an imbalance 
in the population of the K and \ok valleys. Thus, 
the concept of VCD has become a central ingredient for valleytronics applications involving TMDs \cite{mak_lightvalley_2018}, 
and it has provided a strong stimulus for experimental research  
\cite{mai_many-body_2014,wu_electrical_2013,zhang_electrically_2014,scuri_electrically_2020}. 

The emergence of VCD has been ascribed to the lack of an 
inversion center in the hexagonal lattice of TMDs monolayers
\cite{yao_valley-dependent_2008,xiao_coupled_2012}, 
which leads to finite and opposite Berry curvatures 
and orbital angular momenta at the K and \ok valleys.
Earlier theoretical studies of circular dichroism rely on 
the single-particle picture, whereby the ties between dichroic 
absorption, Berry curvature, and orbital magnetization can be 
rigorously established 
\cite{PhysRevLett.68.1943,Altarelli,oppeneer_magneto-optical_1998,souza_dichroic_2008,PhysRevB.75.195121,Vanderbilt:2622531,
schuler_local_2020,PhysRevResearch.2.023139}. 
Despite the remarkable progress in understanding the 
interplay between topological properties and circular dichroism, 
the single-particle picture is unsuitable to account for the 
formation of valley excitons -- which are inherently two-particle excitations --, as 
well as for the non-trivial response of valley excitons to external perturbations 
reported in recent experimental investigations.   
The presence of an external magnetic field leads to 
the valley Zeeman effect, whereby the degeneracy of 
valley excitons is lifted and the ensuing energy 
shift is linear in the field intensity \cite{mak_valley_2014,huang_robust_2020}. 
In presence of an in-plane electric field, valley excitons are subject to 
the exciton Hall effect, a drift velocity transversal to field orientation which 
is reminiscent of the anomalous Hall effect \cite{onga_exciton_2017,kozin_anomalous_2021}. 
Overall, these findings suggest that valley excitons may possess an additional 
orbital degree of freedom governing its interaction with 
external perturbation and their topological properties. 

In this manuscript, we formulate an {\it ab-initio} many-body theory of valley 
excitons and valley-selective circular dichroism based on 
the Bethe-Salpeter equation (BSE). 
Our approach provides a new route to accurately predict the 
degree of valley polarization upon absorption of circularly 
polarized light in an interacting electron-hole gas. 
It is further shown that valley excitons formed in TMDs monolayers 
at either the K or \ok valley upon VCD exhibit a non-trivial chirality. 
In particular, valley excitons are characterized by a finite orbital 
angular momentum, which is inherited by the underlying band structure, 
and provides a new rationale to explain the emergence of VCD in 
interacting electron systems. 
The OAM confers excitons a finite magnetic 
moment that, in concomitance with an external magnetic field, 
underpins a Zeeman-like splitting of the excitonic peaks. 

\section{Results}

\begin{figure*}[t]
\begin{center}
   \includegraphics[width=0.98\textwidth]{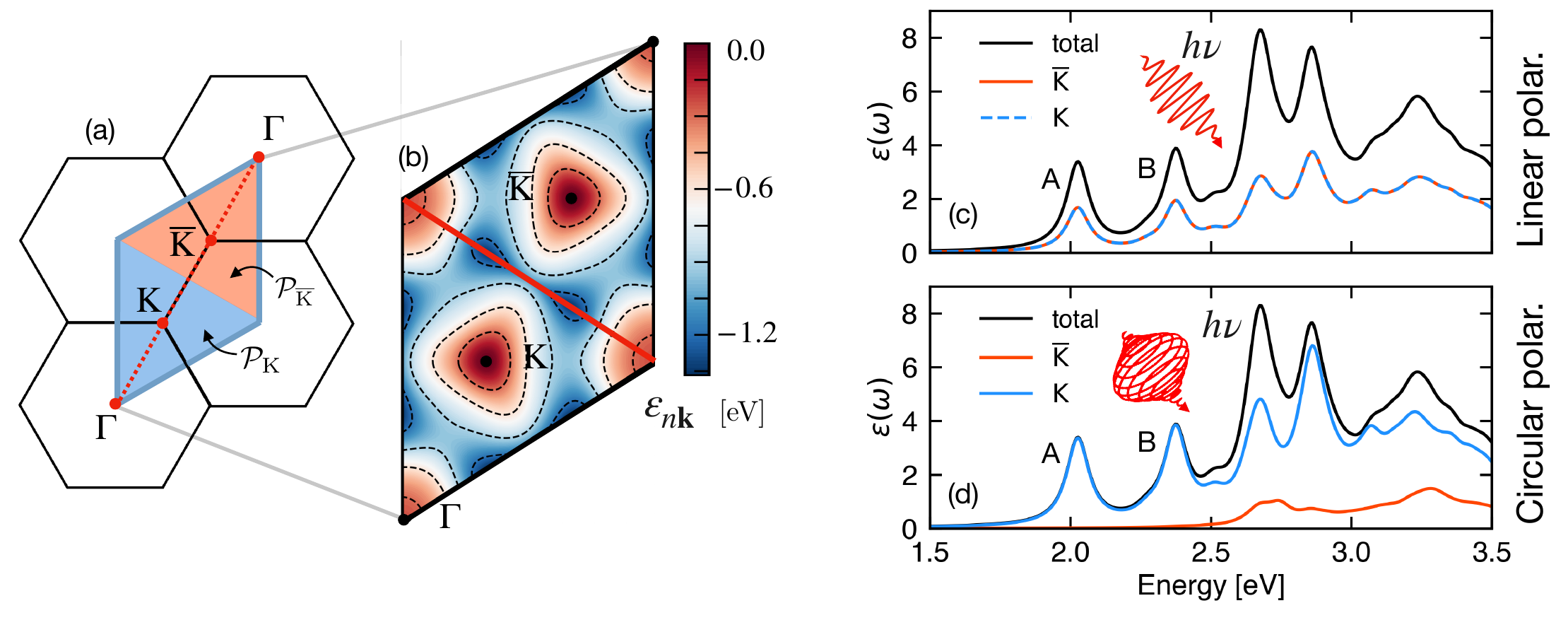}
\caption{\label{fig:bz}
(a) Brillouin zone  and high-symmetry points of monolayer WS$_2$. 
The blue and orange shadings mark the 
$\mathcal{P}_{\rm K}$ and $\mathcal{P}_{\oke}$ regions in the BZ, respectively.
The path $\Gamma$-K-M-\ok-$\Gamma$ is shown as a dotted red line.  
(b) Energy (relative to the valence-band top) of the upper valence 
band for momenta in the rhomboidal BZ. 
The red continuous line delimits the two inequivalent 
regions of the BZ, containing the K and $\overline{\rm K}$ valleys. 
(c-d) Total absorption spectrum  as obtained from the solution of the BSE 
(black), and contribution of the inequivalent K (blue)  and \ok (orange) 
valleys to the absorption for the cases of linear (c) and circular light 
polarization (d) with left-handed chirality. A and B denote the lower and 
higher-energy excitonic peaks, respectively. 
}
\end{center}
\end{figure*}

For definiteness, we focus in the following on monolayer WS$_2$, 
although similar conclusions can be drawn also for WSe$_2$, MoS$_2$, and  MoSe$_2$. 
Monolayer WS$_2$ is a direct band-gap semiconductor \cite{WS2_gap_exp} and its 
valence band is characterized by two degenerate maxima at K
and $\overline{\rm K}$. The hexagonal BZ and the energy of the upper valence band are
illustrated in Fig.~\ref{fig:bz}~(a) and (b), respectively.
To investigate the influence of light polarization on the 
bound excitons formed at the K and $\overline{\rm K}$ valleys, 
we solve the many-body BSE, 
and consider the imaginary part $\ve_2$ of the transverse dielectric
function $\ve= \ve_1 + i \ve_2$ \cite{rohlf_louie_2000}:
\begin{align}
\label{eq:eps1}
\ve_2 (\omega) =  \frac{4\pi^2e^2}{m_e^2 \Omega N_k}
\sum_\lambda \left|\hat {\bm{\epsilon}}\cdot \mathbf{t}^\lambda \right|^2
 \delta   ( E^\lambda  -\hbar \omega) \quad.
\end{align}
Here, $\hat {\bm{\epsilon}}$ denotes the light-polarization unitary vector, $\Omega$  the unit cell  volume, and 
$N_k$ the number of k-points. 
The transition coefficients $ \mathbf{t}_\lambda$ are
defined  as: 
\begin{equation}\label{eq:trans_coeff}
   \bt^\lambda =\sum_{vc}\sum_{\bk}^{\rm BZ} A^\lambda_{vc\bk} \frac{  \langle \psi_{v{\bk}} | \hat {\bf p} | \psi_{c{\bk}} \rangle }{ \ve_{c{\bk}} - \ve_{v{\bk}}  } \quad,
\end{equation}
where $ \hat {\bf p}$ is the momentum operator,  $\psi_{n{\bk}}$ and $\ve_{n{\bk}}$ are single-particle Bloch orbitals and energies,
respectively,
and the sum over $v$ ($c$) runs over 
the valence (conduction) bands.
$E^\lambda$ and $A^\lambda_{vc\bk}$ are the eigenvalues and eigenvectors\cite{rohlf_louie_2000}, respectively, obtained
from the diagonalization of the two-particle BSE Hamiltonian
 $   \sum_{v'c'\bk'}H_{vc\bk,v'c'\bk'}A^\lambda_{v'c'\bk'}=E^\lambda A^\lambda_{vc\bk} $. 

The dielectric function evaluated from 
Eq.~\eqref{eq:eps1} is illustrated in Fig.~\ref{fig:bz}~(c-d) in black. 
As long as the {\it total} absorption is considered 
-- i.e., electronic transitions in the whole BZ 
(as opposed to the valley-dependent absorption within 
a specific valley) -- the dielectric function of WS$_2$ 
is independent of the light-polarization vector~$\hat {\bm{\epsilon}}$. 
The absorption onset 
is dominated by strongly-bound excitons 
localized at K~and~\ok, marked as A and B in Fig.~\ref{fig:bz}~(c) and (d),
in good agreement with earlier calculations and experiments \cite{soc_mos2,zhu_exciton_2015}. 

To investigate the role played by the inequivalent 
K and \ok valleys in the absorption of polarized light, 
we partition the BZ into two regions $\mathcal{P}_{\rm K}$ 
and  $\mathcal{P}_{\oke}$, shaded in Fig.~\ref{fig:bz}~(a), which enclose 
K and \ok, respectively. 
Next, we consider the valley-resolved transition coefficients
$\mathbf{t}^\lambda_{\rm K}$ ($\mathbf{t}^\lambda_{\oke}$), defined 
by restricting the sum over momenta in Eq.~\eqref{eq:trans_coeff} 
to the $\mathcal{P}_{\rm K}$ ($\mathcal{P}_{\oke}$) region in reciprocal space, (i.e.,
replacing $\sum_\bk^{\rm BZ}$  with $\sum_\bk^{\mathcal{P}_{\rm K}}$). 
Additivity follows straightforwardly from these definitions $(\mathbf{t}^\lambda = \mathbf{t}^\lambda_{\rm K} + \mathbf{t}^\lambda_{\oke})$.
We define the valley-resolved dielectric function $\ve_{2, \rm K}$ ($\ve_{2,\oke}$)
by retaining only the valley-resolved transition
coefficients in the evaluation of Eq.~\eqref{eq:eps1}.
In short, $\ve_{2,\rm K}$ ($\ve_{2,\oke}$) accounts for absorption processes resulting 
exclusively from the excitation of electron-hole pairs within the K (\ok) valley, and to 
a good approximation the identity $\ve_2 (\omega)\approx\ve_{2,\rm K} (\omega)+\ve_{2,\oke} (\omega)$ 
holds for frequencies in the vicinity of the absorption onset (see Supplementary Fig.~S2 \cite{sup}). 
\begin{figure}[t]
\begin{center}
   \includegraphics[width=0.65\textwidth]{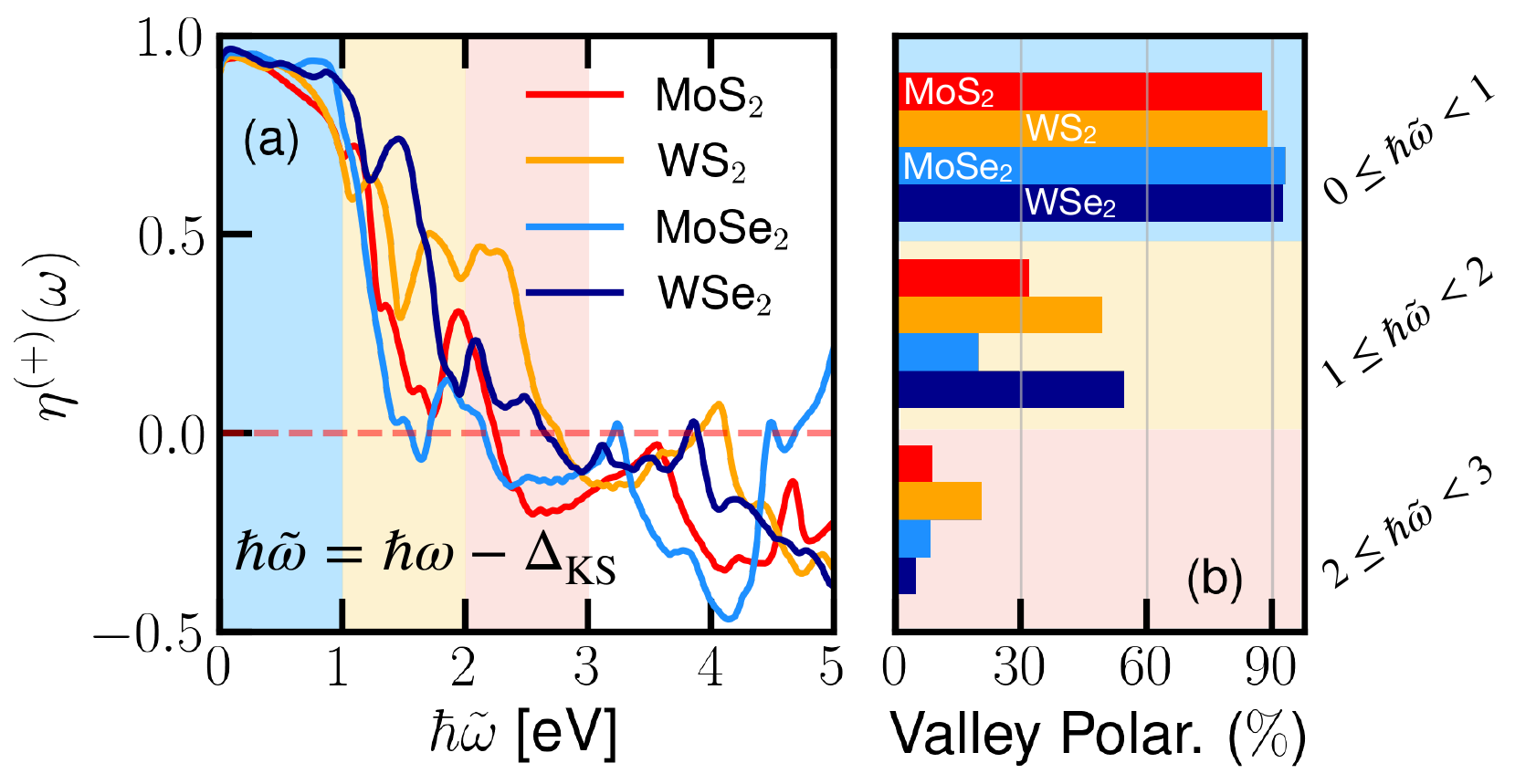}
\caption{\label{fig:eta}
(a)  Degree of valley polarization $\eta^{(+)}$ induced by left-handed circularly-polarized light  in WS$_2$, 
MoS$_2$, MoSe$_2$, and WSe$_2$. 
Energies are relative to the Kohn-Sham band gap $\Delta_{\rm KS}$. 
(b) Average valley polarization for photon energies ranging between 0 and 1 eV (top),  1 and 2 eV (center), 
2 and 3 eV (bottom) above the absorption onset. }
\end{center}
\end{figure}
\begin{figure*}[t]
\begin{center}
   \includegraphics[width=0.98\textwidth]{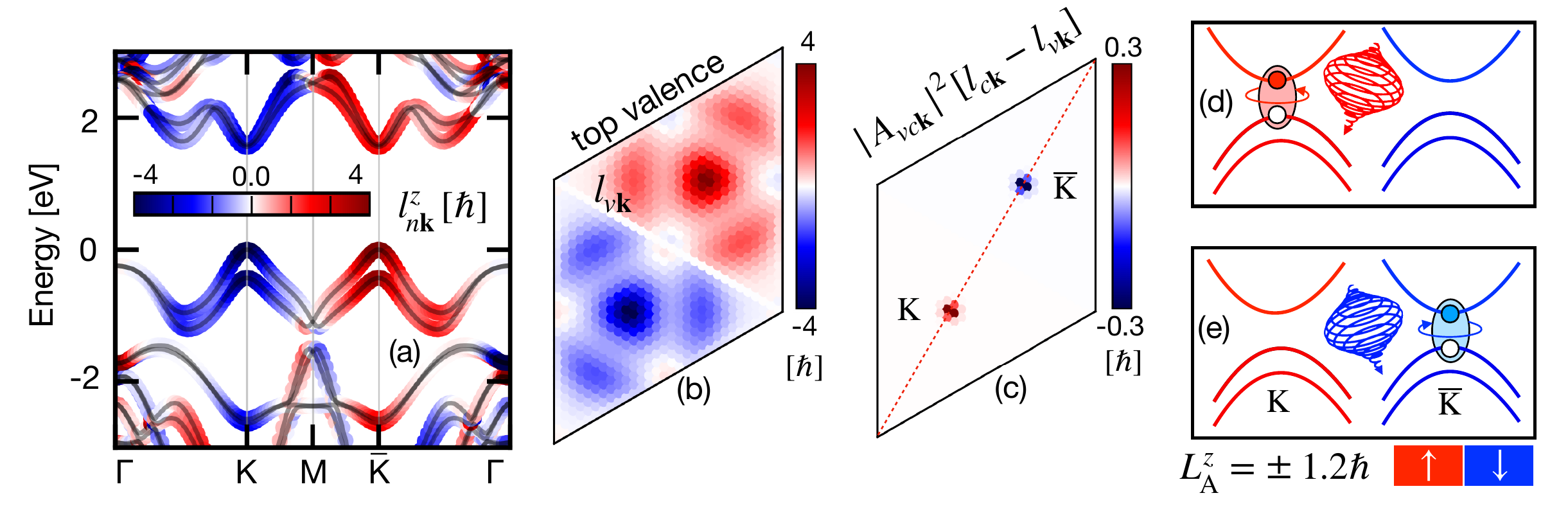}
\caption{\label{fig:orb}
(a) Orbital angular momentum $l^z_{n\bk\sigma}$ superimposed to the 
band structure of monolayer WS$_2$ for momenta along the 
 $\Gamma$-K-M-\ok-$\Gamma$ line. 
(b) OAM for the top valence band for momenta in the BZ. 
(c) Momentum-resolved contribution to the OAM of valley excitons. 
(d) Schematic illustration of valley-selective circular dichroism 
and chiral valley excitons in the TMDs.
}
\end{center}
\end{figure*}
For left- ($+$) and right-handed ($-$) circular polarization, 
the light-polarization vector $\hat{\bm { \epsilon}}$ can be expressed as \cite{Jackson:100964}:
${\bm {\hat\epsilon}_{\pm}} = ({\hat{\bf x} \pm i \hat {\bf y} })/ {\sqrt {2} }$,
where $\hat {\bf x}$ and $\hat {\bf y}$ are Cartesian unit vectors.
By accounting explicitly for the circular polarization via the unitary vectors ${\bm {\hat\epsilon}}_{\pm}$, 
one promptly obtains an explicit expression for the imaginary part of the transverse dielectric function 
at K for circularly-polarized light:
\begin{align}\label{eq:valley_eps}
\ve^{(\pm)}_{2,\rm K}(\omega) &= \frac{1}{2} \left[\xi^\text{K}_{xx}(\omega) + \xi^\text{K}_{yy}(\omega)\right] \mp {\rm Im}\,\left[\xi^\text{K}_{xy}(\omega)\right] \quad,
\end{align} 
where we introduced the {\it dichroic tensor}:
 $   \xi_{\rm K}^{\alpha\beta}(\omega) = \frac{4\pi^2 e^2}{m_e^2 \Omega N_k}\sum_\lambda \left(t_{\rm K}^{\lambda,\alpha}\right)^*t_{\rm K}^{\lambda,\beta}\,\delta(E^\lambda -  \hbar\omega ) \quad.$
The corresponding expression for \ok is obtained by replacing K$\rightarrow\oke$.
A detailed discussion of these expression 
is included in the SI \cite{sup}. 

The emergence of valley-selective circular dichroism  
can be quantified by introducing  
the differential dichroic absorption: 
\begin{align}\label{eq:dic}
D_{\rm K}(\omega) = \ve^{(+)}_{2,\rm K}(\omega) - \ve^{(-)}_{2,\rm K}(\omega) = - 2  {\rm Im}\,[ \xi_{\rm K}^{xy}(\omega)].
\end{align}
In monolayer WS$_2$,  
the {\it total} off-diagonal components of the dichroic 
tensor vanish  at all frequencies ($\xi^{xy}(\omega) =\xi^{xy}_{\rm K}(\omega) + \xi^{xy}_{\oke}(\omega) =  0$), 
leading in turn to a vanishing differential dichroic absorption ($D(\omega)=D_{\rm K}+D_{\rm \overline{K}}= 0$). 
The total absorption spectrum is thus ultimately independent of the helicity of light polarization. 
Conversely, the valley-resolved components of the dichroic tensor 
are finite and opposite in sign at K and \ok (${\rm Im}\,[\xi_{\rm K}^{xy}]= -{\rm Im}\,[\xi_{\rm \oke}^{xy}] \neq 0$)
indicating that, despite the total vanishing dichroism, 
the individual valleys are characterized by a 
non-trivial chiral character, leading to a non-vanishing differential dichroic absorption. 

To analyze these phenomena on a quantitative ground, 
we proceed to investigate the optical response of 
the K and \ok valleys to light with different polarization states. 
For {\it linear} polarization, we consider for definiteness 
a polarization vector $\hat{\bm{\epsilon}}=\hat{\bf x}$.
The absorption in the K and \ok valleys is thus described by the 
$xx$ component of the corresponding valley-resolved dielectric tensors, i.e., 
$\ve^{xx}_{2,\rm K}$ and $\ve^{xx}_{2,\oke}$,  
which are reported in Fig.~\ref{fig:bz}~(c), in orange and blue (dashed), respectively. 
The overlap of $\ve^{xx}_{2,\rm K}$ and $\ve^{xx}_{2,\oke}$ for all 
photon energies reflects the identical response of carriers in the 
K and \ok valleys:  light absorption is equally likely to be mediated 
by optical excitations at K or \ok. 

The situation is qualitatively different when circular polarization is considered.
For left-handed polarization, 
the valley-resolved dielectric functions $\ve^{(+)}_{2,\rm K}$ and $\ve^{(+)}_{2,\oke}$ --  
evaluated from Eq.~\eqref{eq:valley_eps} and illustrated  in Fig.~\ref{fig:bz}~(d) -- 
indicate a strikingly different optical response at K and \ok. 
The absorption onset is dominated by pronounced excitonic peaks 
in the $\mathcal{P}_{\rm K}$ region of the BZ, whereas 
optical excitations in $\mathcal{P}_{\oke}$ are virtually 
suppressed at the absorption onset, giving a sizeable 
contribution only at energies larger than the fundamental gap.
Correspondingly,
light is almost exclusively absorbed in the $\mathcal{P}_{\rm K}$ region of the BZ in the 
vicinity of the absorption onset. 
These findings reflect the formation of valley excitons at K, and the 
absence of excitonic states at \ok. 
The scenario is reversed if the degree of circular polarization is inverted   (not shown). 

To assess the degree of valley selectivity in the 
absorption of polarized light, we define the degree 
of valley polarization $\eta$ for left-handed polarization ($+$),  
\begin{align}
\eta^{(+)}(\omega)  = \frac{ \ve^{(+)}_{2,\rm K}  -\ve^{(+)}_{2,\oke}   } { \ve^{(+)}_{2,\rm K}  + \ve^{(+)}_{2,\oke} } \quad.
\end{align}
For right-handed polarization ($-$) one easily finds $\eta^{(-)}= -\eta^{(+)}$. 
For a given photon frequency $\omega$, $\eta$ assumes values in the range $[-1,1]$, 
where  $\eta(\omega)=0$ denotes absence of valley polarization 
and $\eta(\omega)=\pm 1$ indicates complete valley polarization.
Figure~\ref{fig:eta}~(a) illustrates values of $\eta^{(+)}$ evaluated in the independent 
particle approximation (IPA) for WS$_2$, and the TMDs MoS$_2$, MoSe$_2$, and WSe$_2$ 
in their hexagonal monolayer structure. The valley polarization $\eta$ evaluated 
in the IPA is found to agree well with the result of BSE calculations, as shown in Fig.S3. 
For all compounds, the degree of valley polarization approaches unity in the vicinity of the 
absorption onset, and it decreases rapidly for increasing photon energies. 
Figure~\ref{fig:eta}~(b) reports the average valley polarization for photon 
energies in the vicinity of the absoprtion onset 
(0-1~eV, blue shading in Fig.~\ref{fig:eta}~(a)), and at larger frequencies (1-2~eV in orange shading, and 2-3~eV in red shading).
For WS$_2$, on average valley polarization of 89\%  is obtained 
for energies up to 1~eV above the fundamental gap. 
This value agrees well with the experimental estimate of 84\%
obtained by time-resolved ARPES measurements of the valence band \cite{beyer_80_2019}. 
Valley polarization exceeding 90\% may be obtained in WSe$_2$ and MoSe$_2$. 
These values constitute a theoretical upper limit for 
the degree of valley polarization. 

In the following, we proceed to inspect the 
chiral character of valley excitons via {\it ab-initio}
calculation of the exciton OAM. 
The wave function of an excitonic state $\lambda$ can be expressed in terms of the 
eigenstates of the BSE Hamiltonian $A_{vc\bk}^\lambda$ 
as $\ket{\lambda} = \sum_{vc\bk} A_{vc\bk}^\lambda \ket{ \psi_{v\bk}^{\rm h} \psi_{c\bk}^{\rm e} }$, 
where $\psi_{c\bk}^{\rm e}$ and $\psi_{v\bk}^{\rm h}$ denote 
the Bloch orbitals of electrons and holes, respectively. 
By considering the OAM operator for an 
electron-hole pair $\hat{\bf L} \equiv \hat{\bf l}_{\rm e} +  \hat{\bf l}_{\rm h}$,  
with $\hat{\bf l}_{\rm e(h)} = \hat{\bf r}_{\rm e(h)}  \times \hat{\bf p}_{\rm e(h)}$, 
we obtain an explicit expression for the out-of-plane component of the 
OAM of the exciton $\lambda$: 
\begin{align}\label{eq:Lexc}
{L}^z_\lambda = \bra{\lambda} \hat{ L}^z \ket{\lambda} 
=   \sum_{vc\bk}  |  A_{vc\bk}^\lambda |^2 [ l^z_{c\bk} - l^z_{v\bk}] \quad. 
\end{align}
The derivation of Eq.~\eqref{eq:Lexc} is discussed in detail in the SI \cite{sup}. 
$l^z_{n\bk}$ denotes the single-particle OAM
for a Bloch state $\psi_{n\bk}$, 
and it is given by \cite{oppeneer_magneto-optical_1998}:  
\begin{align}\label{eq:l}
l^z_{n{\bk}} = \frac{2\hbar}{m_e} \sum_{m\neq n}\frac{ {\rm Im}\, [M^{x}_{nm} M^{y}_{mn } ] }{\ve_{m\bk} - \ve_{n\bk} }\quad,
\end{align}
with the abbreviation 
$M^{\alpha}_{nm} =  \langle \psi_{n{\bk}} | \hat { p}^\alpha | \psi_{m{\bk}} \rangle $. 
The single-particle OAM, evaluated from Eq.~\eqref{eq:l}, 
is shown in Fig.~\ref{fig:orb}(a) as a color code
superimposed to the band structure of WS$_2$. 
The lower and upper valence bands exhibit the largest OAM, with opposite sign 
at the K and \ok points. The single-particle OAM of the top valence band 
is further illustrated in Fig.~\ref{fig:orb}(b). 
The total OAM of each valley, obtained from
$l^z_{\rm K}=\int _{\mathcal{P}_{\rm K}}\,l_{v\bk} \,d\bk$, yields $l^z = \pm0.7 \, \hbar$.

These considerations indicate that excitons 
inherit the OAM from the underlying band structure. 
Upon absorption of linearly polarized light, however, 
the exciton OAM vanishes identically owing to the 
compensating contribution from K and \ok. To illustrate this point, we  
report in Fig.~\ref{fig:orb}(c) the momentum-resolved contribution to the OAM of the A
exciton, obtained from the expression $|  A_{vc\bk}^\lambda |^2 [ l^z_{c\bk} - l^z_{v\bk}]$
(see also Eq.~\eqref{eq:Lexc}). 
Conversely, in presence of circularly-polarized  light, 
excitons are localized at either K or \ok, no 
compensation occurs, and {\it chiral excitons} characterized by a finite
OAM can emerge. 
More precisely, the prerequisite for the emergence of chiral excitons is the OAM of the valence and conduction manifold to differ
for band indices $c,v$ and momenta $\bk$ 
contributing to the exciton formation (that is, $l_{c\bk} \neq l_{v{\bk}}$ for 
 $ A_{vc\bk}^\lambda \neq 0$). 
This condition is satisfied by valley excitons 
localized exclusively at K or \ok. 
Evaluation of Eq.~\eqref{eq:Lexc} yields 
$L^z_{\rm A} = \pm1.2~\hbar$ for $A$ valley excitons at ${\rm K}$ and 
\ok, whereas for the $B$ exciton we obtain $L^z_{\rm B} = \pm1.0~\hbar$. 
Because valley excitons are formed by electron-hole pairs in the vicinity of 
K (\ok) 
to a good approximation one has  $L^z_{\lambda {\rm K}} \simeq l_{c{\rm K}} - l_{v{\rm K}}$. The minus sign
reflects the momentum reversal for holes (see SI \cite{sup}). 

The concept of OAM provides a valuable tool to unravel the 
microscopic origin of valley dichroism in TMDs. 
In the independent-particle approximation, the dichroic sum rule 
\cite{oppeneer_magneto-optical_1998,souza_dichroic_2008}
indicates that circular dichroism is a manifestation of 
the finite OAM of independent electrons in the sample. 
While we were unable to derive an exact sum rule relating the 
exciton OAM and the differential dichroic absorption  
in the interacting picture, 
a simple relation between these quantities can be established 
for a two-level system (see SI \cite{sup}), 
suggesting that valley dichroism is a direct manifestation of the exciton chirality. 

\begin{figure}[t]
\begin{center}
   \includegraphics[width=0.68\textwidth]{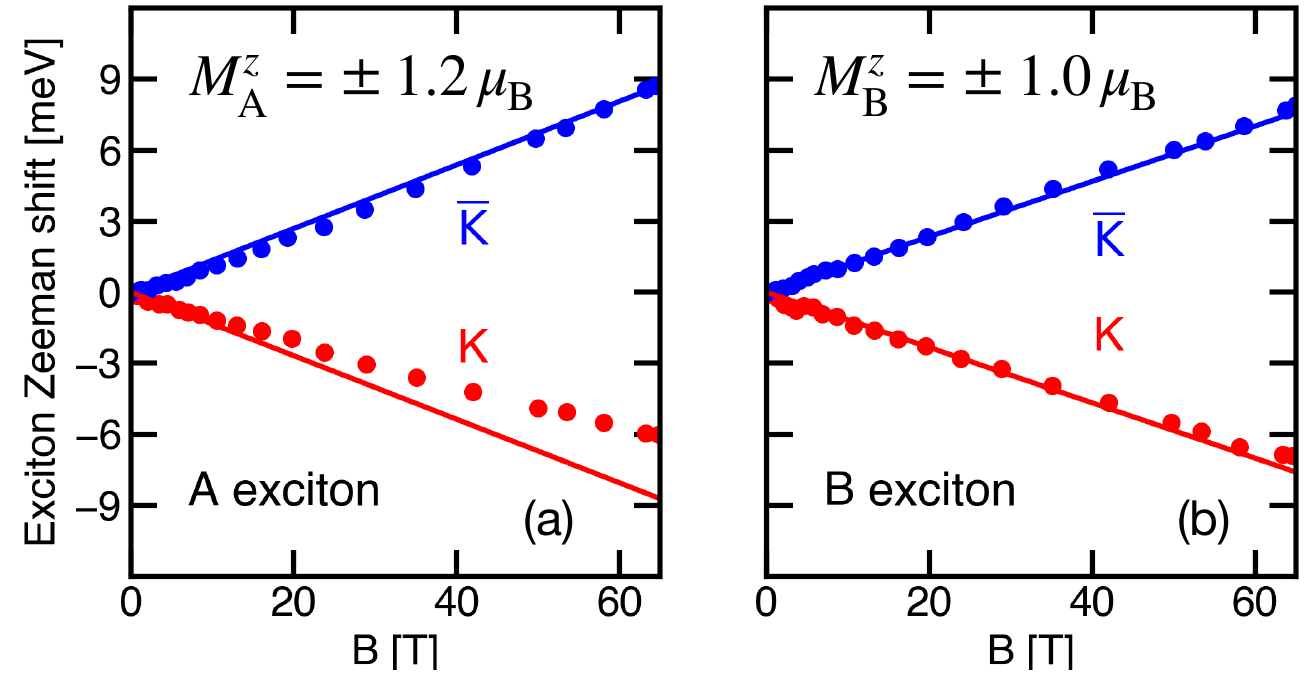}
\caption{\label{fig:zee}
{\it Ab-initio} exciton Zeeman shifts for the A (a) and B excitons (b)  
as a function of external magnetic field (lines). 
Experimental data (dots) are reproduced from Ref.~\cite{stier_exciton_2016}. 
}
\end{center}
\end{figure}

The presence of a finite OAM further confers excitons an orbital magnetic moment 
$M^z_\lambda =- (e/2m_e) L^z_\lambda$, which in turn provides a
route for the interaction of excitons with external magnetic fields
and other spin-orbital degrees of freedom. 
The presence of an external magnetic field in concomitance with 
finite and opposite orbital magnetic moment $M^z_\lambda$
is expected to lift the degeneracy of valley excitons at K and \ok, 
leading to a Zeeman shift of their energy according  to
$\Delta E_b = - {{\bf M}_\lambda \cdot {\bf B}}$. 
The shift of exciton peaks in the absorption spectrum of TMDs 
in presence of external magnetic fields has been observed in 
recent experimental magneto-optical studies \cite{li_valley_2014,srivastava_valley_2015,stier_exciton_2016,li_enhanced_2020,li_optical_2021}. 
We illustrate in Fig.~\ref{fig:zee} 
{\it ab-initio} exciton Zeeman shifts $\Delta E_b$ for WS$_2$ as a function of external fields (lines), 
alongside with magneto-reflectance spectroscopy data from Ref.~\cite{stier_exciton_2016} (dots).  
The excellent agreement between theory and experimental data in Fig.~\ref{fig:zee} corroborates our findings and 
it provides further evidence in support of the orbital degree of freedom of 
valley excitons in transition-metal dichalcogenides. 
These results further suggests that photo-luminescence 
magneto-reflectance spectroscopy constitutes a suitable 
tool to directly probe the OAM and orbital magnetic moment of excitons. 

\section{Conclusions}

In conclusion, we presented a first-principles theory of 
valley-selective circular dichroism and valley excitons 
in monolayer WS$_2$ based on many-body perturbation theory 
and the BSE. 
We showed that valley excitons formed upon absorption of 
circularly-polarized light and localized at either the K or \ok 
valley in the BZ, are chiral quasiparticles characterized by finite
orbital angular momentum and orbital magnetic moment.  
This picture is validated via {\it ab-initio} calculations of the 
exciton Zeeman splitting in good agreement with 
recent experimental studies. 
More generally, the orbital angular momentum 
constitutes an internal degree of freedom of valley 
excitons that influences selections rules, the 
coupling to external perturbations and
other spin-orbital degrees of freedom, and it provides a 
promising concept to unravel the recently discovered 
topological behaviour of excitons \cite{onga_exciton_2017,kozin_anomalous_2021}. 

\section{Methods}
{
\subsection{Computational details}
Optical properties have been computed using the full-potential all-electron 
code {\tt exciting}\cite{exciting_ref,Vorwerk_2019}. For both species, muffin 
tin radii of 2.2\,Bohr have been used with a basis set cut-off 
of $R_{\rm MT}|\bG + \bk|_{\rm max}=8.0$. A grid of $30\!\times\!30\!\times\!1$ $\bk$-points 
was employed for the calculations, local-field effects were included up to a cut-off 
of $|\bG + \bq|_{\rm max}=3.0\,{\rm Bohr}^{-1}.$ The PBE exchange-correlation 
functional\cite{pbe_func} was used to compute the Kohn-Sham states, 100 
unoccupied states were included in the calculation of the dielectric matrix, 
spin-orbit coupling was accounted for through the second variation approach. 
Four valence and four conduction bands have been considered in the solution 
of the BSE, and a broadening of 50\,meV has been applied in the computation 
of the dielectric function.  
The orbital magnetization has been computed using a modified 
version of the {\tt epsilon.x} program, part of  {\tt Quantum Espresso}. 
}

\section{Acknowledgments} 
This project has been funded by the
Deutsche Forschungsgemeinschaft (DFG) -- project numbers 443988403, 424709454, and  182087777. 
Discussions with Michael Bauer are gratefully acknowledged.
MS acknowledges support by the IMPRS for Elementary Processes in Physical Chemistry. 

\providecommand{\latin}[1]{#1}
\makeatletter
\providecommand{\doi}
  {\begingroup\let\do\@makeother\dospecials
  \catcode`\{=1 \catcode`\}=2 \doi@aux}
\providecommand{\doi@aux}[1]{\endgroup\texttt{#1}}
\makeatother
\providecommand*\mcitethebibliography{\thebibliography}
\csname @ifundefined\endcsname{endmcitethebibliography}
  {\let\endmcitethebibliography\endthebibliography}{}

\end{document}